\documentclass[a4paper, amsfonts, amssymb, amsmath, reprint, showkeys, nofootinbib,superscriptaddress, twoside]{revtex4-1}
\usepackage[english]{babel}
\usepackage[utf8]{inputenc}
\usepackage[colorinlistoftodos, color=green!40, prependcaption]{todonotes}
\usepackage{tabularx}
\usepackage{amsthm}
\usepackage{mathtools}
\usepackage{physics}
\usepackage{xcolor, color}
\usepackage{graphicx}
\usepackage[left=23mm,right=13mm,top=35mm,columnsep=15pt]{geometry} 
\usepackage{adjustbox}
\usepackage{placeins}
\usepackage[T1]{fontenc}
\usepackage{lipsum}
\usepackage{csquotes}
\usepackage{svg}
\usepackage[pdftex, pdftitle={Article}, pdfauthor={Author}]{hyperref} 
\usepackage{soul}
\usepackage{multirow}

\newcommand{\expmagicpi}{$473.371(6)$}
\newcommand{\expmagicsig}{$473.117(15)$}

\begin{document}

\title{Magic wavelengths of the Sr (\texorpdfstring{$5s^2\;^1\!S_0 - 5s5p\;^3\!P_1$}{Lg}) intercombination transition near the \texorpdfstring{$5s5p\;^3\!P_1 - 5p^2\;^3\!P_2$}{Lg} transition}

\author{Grady Kestler}
    \affiliation{Department of Physics and Astronomy, University of California, San Diego, California 92093, USA}
    
\author{Khang Ton}
    \affiliation{Department of Physics and Astronomy, University of California, San Diego, California 92093, USA}
    
\author{Dmytro Filin}
    \affiliation{Department of Physics and Astronomy, University of Delaware, Newark, Delaware 19716, USA}
    
\author{Marianna S. Safronova}
    \affiliation{Department of Physics and Astronomy, University of Delaware, Newark, Delaware 19716, USA}
    
\author{Julio T. Barreiro}
    \affiliation{Department of Physics and Astronomy, University of California, San Diego, California 92093, USA}
    
\begin{abstract}

Predicting magic wavelengths accurately requires precise knowledge of electric-dipole matrix elements of nearby atomic transitions. As a result, measurements of magic wavelengths allow us to test theoretical predictions for the matrix elements that frequently can not be probed by any other methods. Here, we calculate and measure a magic wavelength near $473$ nm of the $5s^2\,^1\!S_0 - 5s5p\,^3\!P_1$ intercombination transition of ${}^{88}$Sr. Experimentally, we find \expmagicpi~nm for $\Delta m=0$ ($\pi$-transition) and \expmagicsig~nm for $\Delta m=-1$ ($\sigma^{-}$-transition). Theoretical calculations yield $473.375(22)$~nm  and $473.145(20)$~nm, respectively. Determining magic wavelengths nearby the dominant $461$ nm probe transition of ${}^{88}$Sr holds promise for state insensitive interfacing of strontium atoms and nanophotonic devices with a narrow band of operating wavelengths. Furthermore, the $^3\!P_1$ polarizability is dominated by the contributions to the  $5p^2\, ^3\!P$ levels and excellent agreement of theory and experiment validates both theoretical values of these matrix elements and estimates of their uncertainties. 

\end{abstract}

\maketitle

\section{Introduction}

Laser cooling and probing of atoms trapped in optical potentials, such as in optical dipole traps (ODTs), have been powerful tools for quantum science and technology, e.g., for optical atomic clocks \cite{Ye1999, Bloom2014}. These tools rely on lasers driving an atomic optical transition between two electronic energy levels. In a dipole trap, these two levels have different atomic polarizabilities, resulting in different AC Stark energy level shifts as well as level-dependent optical forces. Consequently, the effectiveness of laser cooling and probing depends on the location of the atoms in the heterogeneous intensity of the optical fields. 

Depending on the atomic electronic structure, it is possible to overcome such challenges by tuning the ODT beam to so-called {\em magic} wavelengths, inducing the same atomic polarizability in each of the two levels \cite{Ido2003, Takamoto2003, Cooper2018}. In the context of probing optical transitions of trapped atoms, magic-wavelength traps have enabled many improvements such as precision spectroscopy \cite{Ye2008, Okaba2014}, the most accurate and stable optical atomic clocks to date \cite{Ludlow2006, Boyd2007, Bloom2014}, and the reduction of decoherence effects in the quantum gates of a quantum computer based on neutral atoms \cite{Safronova2003, Saffman2005, Saffman2016, Weiss2017}. 

In the context of laser cooling, using magic-wavelength traps enables a more efficient transfer of atoms from a magneto-optical trap (MOT) into optical ``tweezers'' \cite{Hutzler2017, Covey2019}, ODTs in which the trap diameter is of the same order as the trapping wavelength \cite{Kaufman2012, Endres2016, Cooper2018}. These traps operate in a regime of large intensities and, thus, large differential AC Stark shifts which are eliminated by using magic wavelengths.

The same principles and improvements for optical tweezers can be applied to interfacing cold atoms with nanoscale waveguides for enhanced atom-photon interactions \cite{Kim2019, Hutzler2017}. Prominent applications of such systems include cavity quantum electrodynamics \cite{Ye2008, Chang2020} and quantum many-body emulations with tunable interaction strengths \cite{Douglas2015, Hung2016}. In these platforms, the fields inside the nanophotonic waveguides can give rise to high intensity evanescent trapping fields near the free-space surface of the device. Hybrid systems, such as nanotapered optical fiber traps have already benefited from magic wavelengths by increasing their coherence times through the elimination of spatially dependent energy shifts near the fiber surface \cite{Goban2012, Lacroute2012}. Furthermore, unlike free-space trapping methods, a repulsive evanescent field is also necessary to counteract the Casimir Polder potential near nanophotonic surfaces. The attractive light field then requires intensities typically more than twice larger than trapping in free-space, thus giving rise to larger AC Stark shifts for off-magic trapping wavelengths.

Most of the current experiments have successfully interfaced cesium \cite{Vetsch2010, Goban2012, Kim2019} or rubidium atoms \cite{Thompson2013, Tiecke2014, Daly2014} with photonic waveguides. However, the bosonic isotope of strontium $^{88}$Sr has several advantages \cite{Okaba2014}. It is less susceptible to magnetic fields, due to a spherically symmetric ground state and no electronic or nuclear spin. Compared to other atomic species or strontium isotopes, this isotope also has a smaller collisional scattering length, thus minimizing any dephasing effects in matter-wave interferometry. Furthermore, with strontium, the easily achievable single-digit micro-Kelvin temperatures allow us to trap and perform spectroscopy with milliwatts of optical power instead of watts \cite{Zheng2020}, even after accounting for differing contributions to trapping strengths.

An initial step towards efficiently loading $^{88}$Sr into nanophotonic evanescent fields is to identify magic wavelengths of the narrow-line intercombination cooling transition that leads to micro-Kelvin temperatures. However, there is an added complication with photonic devices, e.g., optical ring resonators, whereby they are difficult to design for strongly guided modes across a wide range of wavelengths. Probing trapped $^{88}$Sr atoms from the nanowaveguide with the dominant $5s^2\,^1\!S_0  - 5s5p\,^1\!P_1$ (henceforth $^1\!S_0 - ^1\!P_1$) transition at $\approx$461~nm, drives the material choices of the photonic devices for operation in the blue spectrum, and this in turn the choice of trapping magic wavelengths. Current state insensitive trapping of ${}^{88}$Sr makes use of magic wavelengths at $\approx$515 \cite{Cooper2018} and $\approx$813~nm \cite{Okaba2014} both of which are difficult to accommodate in a device tuned for the blue spectrum.


Here we calculate and experimentally verify magic wavelengths of the $5s^2\,^1\!S_0 - 5s5p\,^3\!P_1$ (henceforth ${^1\!S_0} - {^3\!P_1}$) intercombination transition of $^{88}$Sr.

\section{Theory}

For a linearly polarized laser field, the differential AC Stark shift between a ground and an excited state is proportional to the differential atomic polarizability $\Delta \alpha = (\alpha_{e} - \alpha_{g})$ and the intensity $I$ of the light field, $h\Delta f \propto -\Delta \alpha I$, where $\alpha_{g}$ ($\alpha_{e})$ is the atomic polarizability of the ground (excited) state. The polarizability is further separated into its irreducible parts, the scalar, vector, and tensor polarizabilities ($\alpha_{0}, \alpha_{1}$, and $\alpha_{2}$ respectively). The scalar component acts as a level-independent shift and the tensor component has a quadratic dependence on the $m_{J}$ Zeeman sublevels of the excited state. The vector field is only present with circularly polarized light and behaves in the same capacity as a magnetic Zeeman shift.

The spherically symmetric ground state has zero angular momentum and thus only contains the scalar component. The excited state ${}^{3}\!P_{1}$ does not have vanishing angular momentum and must include a contribution from the tensor polarizability as well. This contribution depends on the Zeeman substate and a geometric term based on the polarization angle relative to the quantization axis $\theta_{p}$. We can neglect the vector term completely by considering pure linear polarization. Then the total polarizability becomes \cite{Zheng2020}
\begin{equation}
    \label{eq:pol}
    \alpha = \alpha_{0} + \alpha_{2}\bigg(\frac{3\cos^{2}\theta_{p} - 1}{2}\bigg)\frac{3m^{2} - J(J+1)}{J(2J-1)}.
\end{equation}
For $\theta_{p}=0$ considered here, $\alpha=\alpha_0-2\alpha_2$ for $m=0$ and $\alpha=\alpha_0+\alpha_2$ for $|m|=1$.
The magic wavelengths are determined as crossing points of the $^1\!S_0$ and $^3\!P_1$ polarizability curves for $|m|=0,\, 1$.

The frequency-dependent scalar polarizability, $\alpha(\omega)$, of an
atom in a state $v$ may be separated into a core polarizability
and a contribution from the valence electrons. There is no core contribution to the tensor polarizability as angular momentum of the core is zero. The core polarizability is a sum of the
polarizability of the ionic Sr$^{2+}$ core and a ``vc'' term that
compensates for a Pauli principle violating core-valence excitation from the core to the valence shells.
 The core polarizability is small and essentially independent from frequency in the range of interest for this paper, so it is sufficient to use its static value
calculated
in the random-phase approximation (RPA)  \cite{Sr2013}.

\begin{table} [t]
\caption{\label{tab1t} Contributions to the $5s^2\,^1\!S_0$ polarizability in atomic units at the 473.1445- and  473.375-nm $^1\!S_0-^3\!P_1$ magic wavelengths.
Transition energies $\Delta E$ \cite{NIST} in cm$^{-1}$ and the value of the electric-dipole matrix element $D$ \cite{yasuda06,HeiParSan20}
(in atomic units) are also listed.}
\begin{ruledtabular}
\begin{tabular}{lcccc}
\multicolumn{1}{c}{Contr.}& \multicolumn{1}{c}{$\Delta E$}&  \multicolumn{1}{c}{$D$}& \multicolumn{1}{c}{$\alpha_0$}&  \multicolumn{1}{c}{$\alpha_0$} \\
\multicolumn{3}{c}{}& \multicolumn{1}{r}{473.1445~nm}&
  \multicolumn{1}{r}{473.375~nm} \\  
\hline
$5s5p$\,$^1\!P_1$  & 21698 &  5.248(12)&   3624(17)&   3560(16) \\
Other            &       &           &        8  &        8   \\
Core+vc          &       &           &        5  &        5   \\
Total            &       &           &  3637(17) & 3573(16)  \\
  \end{tabular}
\end{ruledtabular} \end{table}

\begin{table*} [th]
\caption{\label{tab2t} Contributions to the $5s5p\,^3\!P_1$ polarizability in atomic units at the 473.1445- and  473.375-nm$^1\!S_0 - ^3\!P_1$ magic wavelengths.
Transition energies $\Delta E$ in cm$^{-1}$ \cite{NIST} and the recommended values of the electric-dipole matrix elements $D$ (see the text)
in atomic units are also listed. $5s5p\,^3\!P_1$ $|m|=1$ polarizabilities are obtained as $\alpha(|m|=1)=\alpha_0+\alpha_2$; $m=0$ polarizabilities are obtained as $\alpha(m=0)=\alpha_0-2\alpha_2$.}
\begin{ruledtabular}
\begin{tabular}{lcccccccc}
\multicolumn{1}{c}{Contributions}& \multicolumn{1}{c}{$\Delta E$}&  \multicolumn{1}{c}{$D$}& \multicolumn{1}{c}{$\alpha_0$}&  \multicolumn{1}{c}{$\alpha_2$} &\multicolumn{1}{c}{$\alpha(|m|=1)$}  &
\multicolumn{1}{c}{$\alpha_0$}&  \multicolumn{1}{c}{$\alpha_2$} &\multicolumn{1}{c}{$\alpha(m=0)$}  
\\
\multicolumn{3}{c}{}& \multicolumn{3}{c}{$\lambda=473.1445$~nm}&
  \multicolumn{3}{c}{$\lambda=473.375$~nm} \\ 
  \hline
$5s4d$\,$^3\!D_1$ &  3655  &  2.318(5)  &    -2       &  -1     &   -3     &  -2     &    -1    &     0 \\
$5s4d$\,$^3\!D_2$ &  3714  &  4.013(9)  &    -7       &   1     &   -6     &  -7     &     1    &    -8 \\
$5s6s$\,$^3\!S_1$ &  14534 &  3.435(14) &   -36       & -18     &   -53    &  -36    &   -18    &     0  \\
$5s5d$\,$^3\!D_1$ &  20503 &  2.005(20) &   -153(3)   & -76(2)  &  -229    & -155(3) &   -78(2) &     0  \\
$5s5d$\,$^3\!D_2$ &  20518 &  3.671(36) &   -524(10)  &  52(1)  &  -472    & -533(10)&    53(1) &   -640 \\
$5p^2$\,$^3\!P_0$ &  20689 &  2.658(27) &   -382(8)   &  382(8) &    0     & -391(8) &    391(8)&  -1174 \\
$5p^2$\,$^3\!P_1$ &  20896 &  2.363(24) &   -565(11)  & -283(6) &  -848    & -591(12)&   -296(6)&     0   \\
$5p^2$\,$^3\!P_2$ &  21170 &  2.867(29) &   5714(116) & -571(12)&   5142   & 4420(89)&   -442(9)&   5304  \\
$5p^2$\,$^1\!D_2$ &  22457 &  0.228(11) &     1       &    0    &   1      &    1    &     0    &     1   \\
$5p^2$\,$^1\!S_0$ &  22656 &  0.291(15) &     1       &   -1    &   0      &    1    &    -1    &     4  \\
$5s7s$\,$^3\!S_1$ &  22920 &  0.921(28) &     12(1)   &    6    &   18     &   12(1) &     6    &     0  \\
Other           &        &            &    81(2)    &   1     &  82      &  81(2)  &    1     &    80  \\
Core+vc         &        &            &     6       &         &  6       &  6      &          &    6  \\
Total           &        &            &   4146(117) & -509(15)& 3637(118)& 2805(91)&  -384(14)&  3573(92)
   \end{tabular}
\end{ruledtabular} \end{table*}

We calculate the valence polarizabilities using a hybrid approach that combines configuration iteration
and a linearized coupled-cluster method (CI+ all order) \cite{2009,Sr2013,HeiParSan20}. The valence part of the polarizability for state $v$ with the total angular momentum $J$ and projection
$m$ is determined  by
solving the inhomogeneous equation of perturbation theory in the
valence space, which is approximated as \cite{kozlov99a}
\begin{equation}
(E_v - H_{\textrm{eff}})|\Psi(v,m^{\prime})\rangle = D_{\mathrm{eff},q} |\Psi_0(v,J,m)\rangle.
\label{eq1}
\end{equation}
The parts of the wave function $\Psi(v,m^{\prime})$
with angular momenta of $J^{\prime} = J, J \pm 1$ are then used  to determine
the scalar and tensor polarizabilities. The effective Hamiltonian $H_{\textrm{eff}}$ used in the configuration interaction (CI) calculations includes the all-order corrections calculated using the
linearized coupled-cluster method with single and double excitations.
The effective dipole operator
$D_{\textrm{eff}}$  includes RPA corrections.
This approach automatically includes contributions from all possible states.

The magic wavelengths of interest are very close to the resonances, the wavelength of the $5s5p ^3\!P_1 - 5p^2$\,$^3\!P_{2}$ transition is 472.4 nm, and it is essential that such relevant transition energies are accurate in a polarizability computation. Therefore,  
we extract several contributions
to the valence polarizabilities using the sum-over-states formulas \cite{mitroy10}:
\begin{eqnarray}
    \alpha_{0}^v(\omega)&=&\frac{2}{3(2J+1)}\sum_k\frac{{\left\langle k\left\|D\right\|v\right\rangle}^2(E_k-E_v)}{     (E_k-E_v)^2-\omega^2}, \label{eq-1} \nonumber \\
    \alpha_{2}^v(\omega)&=&-4C\sum_k(-1)^{J+j_k+1}
            \left\{
                    \begin{array}{ccc}
                    J & 1 & j_k \\
                    1 & J & 2 \\
                    \end{array}
            \right\} \nonumber \\
      & &\times \frac{{\left\langle
            k\left\|D\right\|v\right\rangle}^2(E_k-E_v)}{
            (E_k-E_v)^2-\omega^2} \label{eq-pol},
\end{eqnarray}
where a quantity in $\{ \}$ brackets is a Wigner 6-j symbol and $j_k$ as the total angular momentum of the state $k$.
          The value of  $C$ is given by
\begin{equation}
            C =
                \left(\frac{5J(2J-1)}{6(J+1)(2J+1)(2J+3)}\right)^{1/2}. \nonumber
\end{equation}
\begin{figure}[t]
    \centering
    \includegraphics[width=\linewidth]{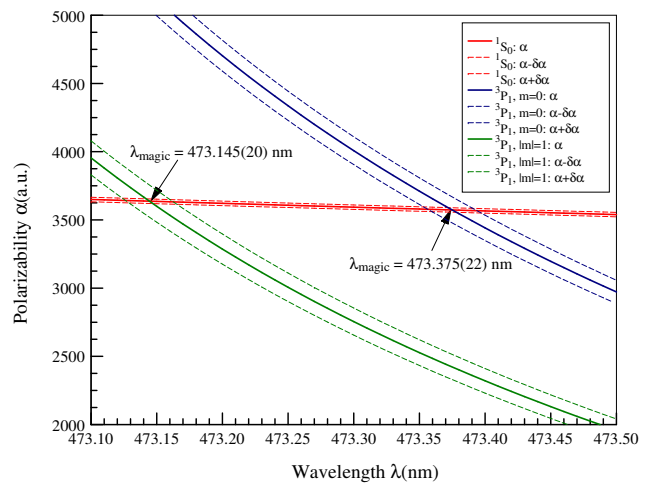}
    \caption{Calculated polarizabilities for $5s^{2}\,^1\!S_{0}$ and $5s5p\,^3\!P_{1}$, $m=0$, and $|m|=1$ levels and corresponding magic wavelengths.}
    \label{fig:pol_calc}
\end{figure}
\noindent We replace such contributions with values obtained using the experimental energies \cite{NIST} and recommended values of the matrix elements from Refs. \cite{yasuda06,Sr2013,Sr2014,HeiParSan20}.
In more detail, we first calculate four such contributions for the $^1\!S_0$ polarizability and 15 contributions for the $^3\!P_1$ polarizability using  \textit{ab initio} energies and matrix elements that exactly correspond to our calculations using the inhomogeneous equation (\ref{eq1}) and determine the remainder contribution of  all other states.  Then, we compute the same  terms  using the experimental energies \cite{NIST} and recommended values of matrix elements from Refs. \cite{yasuda06,Sr2013,Sr2014,HeiParSan20}. Theoretical values of the  matrix elements are retained where recommended values are not available.
The recommended value for the $^1\!S_0-^1\!P_1$ matrix element is from the $^1P_1$ lifetime measurement ~\cite{yasuda06}, but with increased uncertainty based on the discussion in \cite{HeiParSan20}. 
 After this procedure, we add core and the remainder contribution from the other states to these values to obtain the final results.
This calculation is illustrated by Tables ~\ref{tab1t} and \ref{tab2t} where we list contributions to the $^1\!S_0$ and $^3\!P_1$ polarizabilities in atomic units at the 473.1445- and  473.375-nm $^1\!S_0 - ^3\!P_1$ magic wavelengths.
Transition energies $\Delta E$ \cite{NIST} in cm$^{-1}$ and the recommended value of the electric-dipole matrix elements $D$ \cite{yasuda06,Sr2013,Sr2014,HeiParSan20}
(in atomic units) are also listed. The breakdown of the contributions shows that the $^3\!P_1$ polarizability is strongly dominated by the contributions from the  $5s5p \,^3\!P_1 - 5p^2\, ^3\!P_J$ transitions.

The determination of the magic wavelengths and their uncertainties is illustrated in Fig.~\ref{fig:pol_calc},
where we plot the calculated polarizabilities for $5s^{2}\,^1\!S_{0}$ and $5s5p\,^3\!P_{1}$, $|m|=0,\, 1$ states. The magic wavelengths are obtained as the crossing points. We determine the uncertainty of all polarizability contributions from  the estimated uncertainties in the values of the matrix elements, adding all uncertainties in quadrature. We plot $\alpha+\delta \alpha$ and $\alpha-\delta \alpha$  for all levels and determine the uncertainty in the values of the magic wavelength from crossings of these additional curves. The final values are in excellent agreement with the experiment, confirming the central values and the 1\% estimated uncertainties of the  $5s5p \,^3\!P_1 - 5p^2\, ^3\!P_J$ matrix elements.

\section{Experimental Methods}

\begin{figure}[t]
    \centering
    \includegraphics[width=\linewidth]{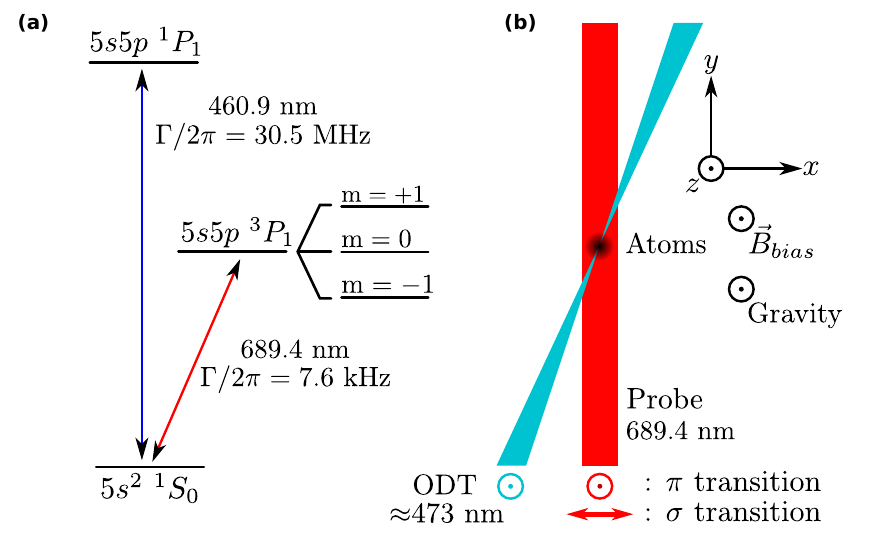}
    \caption{(a) Relevant level diagram and lifetimes for ${}^{88}$Sr. The ``red'' MOT uses the ${}^1\!S_0 - {}^3\!P_1$ narrow-line, intercombination transition, whereas preliminary cooling stages operate on the ${}^1\!S_0 - {}^1\!P_1$ dipole transition. (b) Setup to investigate the magic wavelengths for the ${}^1\!S_0$ and ${}^3\!P_1$ transitions for $m=0$ and $m=-1$ Zeeman sublevels. Atoms in the ``red'' MOT (black ball) are collected in a focused ODT beam, which overlaps with the probe beam. The ODT polarization and $\vec{B}_{\rm{bias}}$ are set to be parallel such that $\theta_{p}=0$. The probe beam polarization is set to vertical for the $m=0$ ($\pi$) measurements and horizontal for the $m=-1$ ($\sigma^{-}$) measurements.}
    \label{fig:dssc_setup}
\end{figure}

Cold atoms are prepared in our experiment \cite{Kestler2019} using established laser cooling techniques for strontium \cite{Campbell2017, Nicholson2015} (see the Appendix). At the end of the narrow-line ``red'' MOT cooling stage, see Fig. \ref{fig:dssc_setup}(a), we capture around $5\times 10^{5}$ atoms at $\approx$1~$\mu$K. The ODT beam is then overlapped with the cold atomic cloud as shown in Fig. \ref{fig:dssc_setup}b. For detection, a 689~nm laser beam is scanned in frequency to probe the ${}^1\!S_0 - {}^3\!P_1$ narrow-line transition for spin-resolved atomic absorption imaging.

\begin{figure}[t!]
    \centering
    \includegraphics[width=\linewidth]{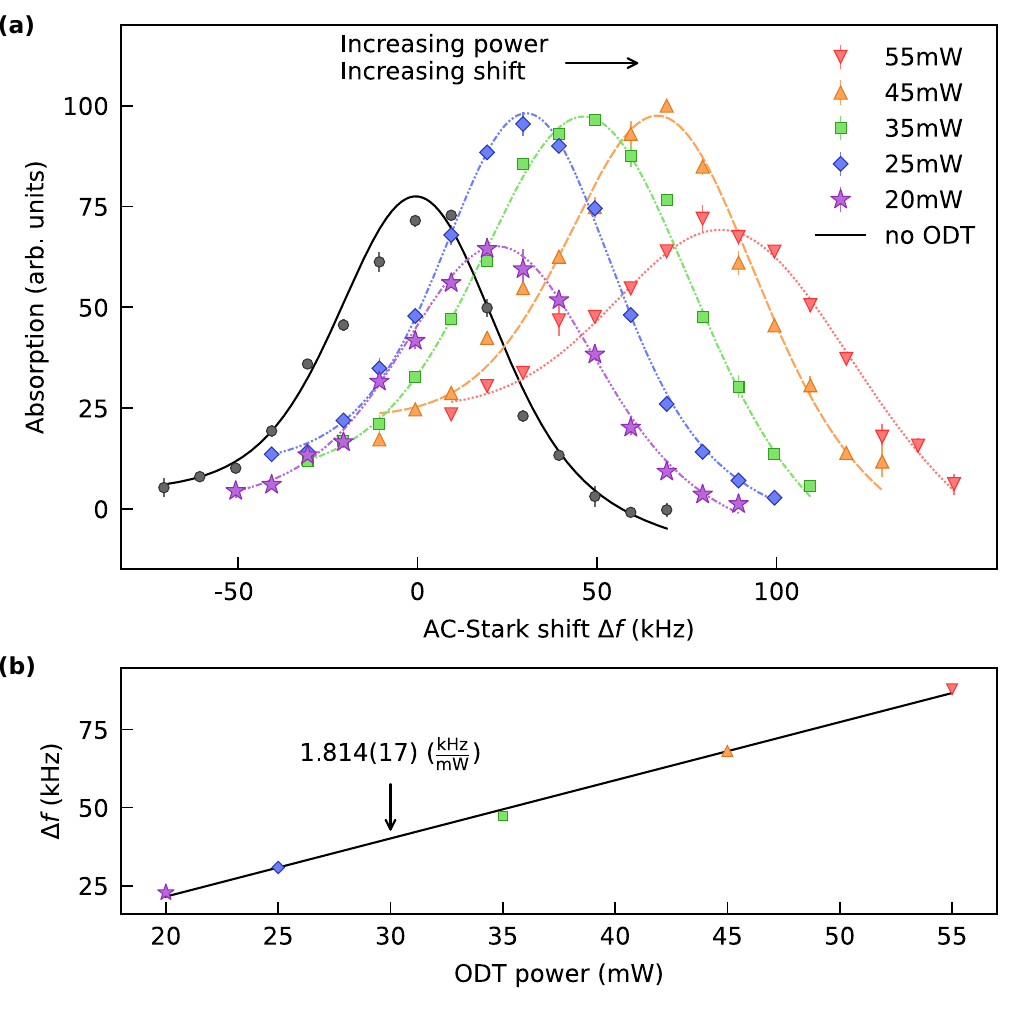}
    \caption{(a) Sample absorption spectroscopy scans for varying powers of an ODT at a wavelength of 473.65~nm for the $\pi$ transition. The AC Stark shift axis is centered with respect to a reference scan without ODT (black solid line). Each scan is fitted to a Voigt plus linear model (see the Appendix). (b) AC Stark shift vs ODT power. At this wavelength, the Voigt centers increase in frequency with the ODT beam power and result in a positive differential Stark shift coefficient (DSSC).} 
    \label{fig:dssc_55}
\end{figure}

The magic wavelength is determined by measuring the differential AC Stark shift at wavelengths both above and below the predicted magic value. First, a reference spectroscopic scan without the ODT is performed at 1~ms after the atoms are released from the ODT. Then, at ODT powers between 20 and 70~mW, corresponding to trap depths between 2~$\mu$K and 10~$\mu$K, we measure the shift in the resonance due to the presence of the ODT beam [Fig. \ref{fig:dssc_55}a]. The slope which characterizes the linear relationship of the detuning and trapping powers is defined as the differential Stark shift coefficient (DSSC) and given by $\Delta f/P$, where $\Delta f$ is the differential AC Stark shift and $P$ is the ODT power [Fig. \ref{fig:dssc_55}b]. For positive (negative) DSSC, the differential AC Stark shift will increase (decrease) with increasing powers and is zero at the magic wavelength.

Spin-resolved spectroscopy is achieved by applying a large bias magnetic field $B_{\rm bias}$ = 200(11)~mG in the $+z$-direction. This separates the excited state Zeeman levels by 420~kHz providing sufficient splitting of the $m=0$ and $m=-1$ resonances, whose widths are primarily limited by Doppler broadening on the order of $30$ kHz. Additionally, the probe beam polarization is set to vertical ($\pi$) or horizontal ($\sigma$) depending on the measurement [Fig. \ref{fig:dssc_setup}(b)]. Since $m=-1$ measurements are sensitive to magnetic fields (2.1~kHz/mG), we perform an additional, second resonance scan after two of the five ODT powers to ensure the bias field is stable.

The ODT polarization is chosen to be vertical and parallel ($\theta_{p}=0$) to the quantization axis defined by the $\vec{B}_{\rm bias}$ consistent with the theoretical predictions. This guarantees the excited state polarizabilities for the Zeeman sublevels are given by $\alpha = \alpha_{0} - 2\alpha_{2}$ for $m=0$ and $\alpha = \alpha_{0} + \alpha_{2}$ for $|m|=1$. Measuring both differential polarizabilities allows us to separate and recombine the scalar and tensor polarizabilities for any $\theta_{p}$ \cite{Zheng2020}.

\section{Results and Discussion}
\label{sec:results}

\begin{figure}[t!]
    \includegraphics[width=\linewidth]{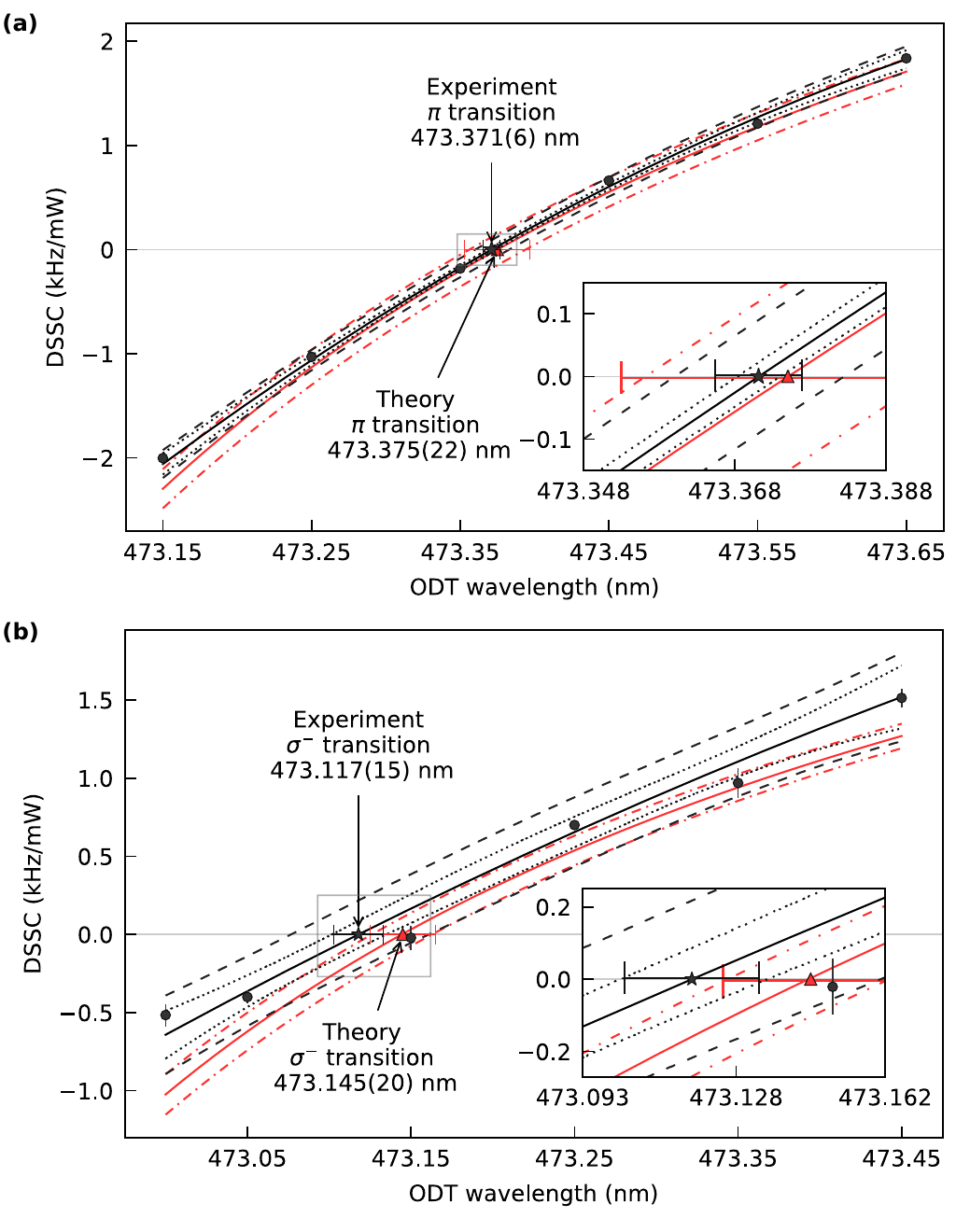}
    \caption{Measured and calculated DSSC as a function of ODT wavelength for (a) $\pi$ and (b) $\sigma^{-}$  $^1\!S_0-^3\!P_1$ transitions. The proximity ($\approx$1~nm) to the $5p^2\, ^3\!P_2$ excited state resonance at 472.4~nm results in a quadratic behavior of the data (circles). Fitting to a quadratic model (solid black line) and employing a Monte Carlo root finding method, we determine the magic wavelengths at the zero crossings, indicated by the black star. The black dotted lines provide the 95\% confidence interval of the fit and the black dashed lines provide the 95\% prediction interval of the data. The red solid line is a theoretical fit to the DSSC measurements that is $\propto {\rm \Delta\alpha}/w^{2}$ where $\Delta\alpha$ is the theoretical differential polarizability calculated from Fig. \ref{fig:pol_calc} and $w$ is a free fit parameter for the ODT beam waist at the atoms. The red triangle indicates the predicted magic wavelength and the red dashed-dot lines correspond to the theoretical $\delta\alpha$ uncertainties in Fig. \ref{fig:pol_calc}.}
    \label{fig:results}
\end{figure}

With the DSSC measurements at each wavelength of Fig. \ref{fig:results}, the data (black) exhibits a quadratic behavior due to the proximity of the nearby excited state resonance at 472.4~nm. Thus, we fit the experimental data to a quadratic model (Fig. \ref{fig:results}). We calculate the magic wavelength at the zero crossing by employing a root finding Monte Carlo approach, which gives \expmagicpi~nm for $m=0$ and \expmagicsig~nm for $m=-1$.

The theoretical calculation of the differential polarizability $\Delta\alpha$ relates to the DSSC by $\Delta f = 1/(2h\epsilon_{0}c) I \Delta\alpha$ where the beam intensity for a Gausssian beam $I=2 P/(\pi w^{2})$ depends on the beam waist at the atoms $w$. In Fig. \ref{fig:results} (red) we fit the theoretical differential polarizability to our data leaving the beam waist as a free parameter. At the magic wavelength, the differential polarizability is zero and independent of beam waist. The fit is consistent with the theoretical prediction, providing a beam waist of $w=44(2)$~$\mu$m for $\pi$ polarization and $w=58(1)$~$\mu$m for $\sigma^{-}$ polarization, similar to out-of-vacuum measurements for each independent experiment. The red dashed-dot lines are theoretical $\delta\alpha$ uncertainties from Fig. \ref{fig:pol_calc}.

There is a systematic error produced by an offset of $\theta_{p}$ from $0$ and is already accounted for in Fig. \ref{fig:results}. We measure $\theta_{p}$ using an indirect method by comparing the excited state populations of the $\pi$ and $\sigma^{-}$ transitions \cite{Zheng2020} (see the Appendix). We find the angle to be $15\pm6^{\circ}$ from $0$ corresponding to a shift in the magic wavelengths of $+10$~pm for the $\pi$ transition and $-15$~pm for the $\sigma^{-}$ transition. A summary of systematic errors and uncertainties are compiled in Table \ref{tab_errs}.

\begin{table}
\caption{\label{tab_errs} Various contributions to the magic wavelength measurement error in picometers. Statistical errors, ODT power fluctuations, and atom count fluctuations are propagated through the fitting procedure and included in the measurement error row. All contributions and corrections are accounted for in Fig. \ref{fig:results}.}
{\renewcommand{\arraystretch}{1.2}
\begin{tabular}{p{2.25cm}>{\centering}p{0.25cm}>{\centering}p{1.15cm}>{\centering}p{1.15cm}>{\centering}p{0.25cm}>{\centering}p{1.15cm}>{\centering\arraybackslash}p{1.15cm}}
&&\multicolumn{2}{c}{Offset}&&\multicolumn{2}{c}{Uncertainty}\\\cline{3-4}\cline{6-7}Contribution&&$\pi$&$\sigma^{-}$&&$\pi$&$\sigma^{-}$\\
\hline
Measurement && 0 & 0 && 4 & 14 \\
$\theta_{p}$ && +10 & -15 && 4 & 6 \\
Wavemeter && +0.4 & +0.4 && 0.1 & 0.1 \\
Total&& +10.4 & -14.6 && 6 & 15 \\
\hline\hline
\end{tabular}}
\end{table}

\section{Conclusion}

In this paper, we report on our investigation of magic wavelengths of the $5s^2\,^1\!S_0 - 5s5p\,^3\!P_1$ intercombination transition of $^{88}$Sr. By inducing a Zeeman splitting of the excited state, we measure one magic wavelength for the $\Delta m=0$ ($\pi$ transition) and another for the $\Delta m=-1$ ($\sigma^{-}$ transition). Precision atomic calculations predicts values of 473.375(22)~nm ($\pi$) and 473.145(20)~nm ($\sigma$) that are consistent with our experimental results of \expmagicpi~nm and \expmagicsig~nm, respectively.
The $^3\!P_1$ polarizability is strongly dominated by the contributions from the  $5s5p \,^3\!P_1 - 5p^2\, ^3\!P_J$ transitions for which there are no prior experimental benchmarks.  As the $^1\!S_0$ polarizability is essentially defined by the experimentally known $^1\!S_0-^1\!P_1$ matrix elements, the magic wavelengths probe $5s5p \,^3\!P_1 - 5p^2\, ^3\!P_J$ transitions. 
Excellent agreement of theory and experiment validates both theoretical values of these matrix elements and estimates of their uncertainties.

Employing narrow-line magic wavelengths eliminates position- and level-dependent forces, allowing for more efficient optical trapping. Additionally, the proximity of an attractive $\approx$473-nm magic trapping wavelength, and a repulsive magic wavelength at $\approx$436~nm, to the ${}^1\!S_0 - {}^1\!P_1$ dominating transition at $\approx$461~nm is key to trapping ${}^{88}$Sr on the evanescent fields of nanoscale waveguides, particularly with a MOT operating with a narrow-linewidth transition. These evanescent fields often give rise to elliptical polarizations which induce a vector shift of the excited state ($3P_1$) polarizabilities, something not considered in this paper. However, depending on the geometry of the waveguides in question, compensation schemes \cite{Lacroute2012} can cancel the elliptical components and the results reported here are then applicable.

\begin{acknowledgements}
We would like to thank P. Lauria for helpful insight, discussions, and assistance building the experimental apparatus. We acknowledge the support of the Office of Naval Research under Grants No.
N00014-20-1-2513 and N00014-20-1-2693 and a UC San Diego Lattimer Faculty Fellowship.
This research was supported in part through the use of University of Delaware HPC Caviness and DARWIN computing systems: DARWIN - A Resource for Computational and Data-intensive Research at the University of Delaware and in the Delaware Region, Rudolf Eigenmann, Benjamin E. Bagozzi, Arthi Jayaraman, William Totten, and Cathy H. Wu, University of Delaware, 2021, \cite{udel}

\end{acknowledgements}

\begin{appendix}

\section*{Appendix}

\subsection{Experimental setup}

In our experiment, strontium atoms are expelled in a collimated atomic beam from an oven heated to 480~C. The atoms, Doppler shifted due to the high velocity, are slowed by a Zeeman slower comprised of a counter propagating laser beam detuned by $\Delta/2\pi = $-540~MHz from the ${}^1\!S_0 - {}^1\!P_1$ transition and a spatially dependent magnetic field designed to keep atoms on resonance as they are slowed. Two dimensional MOTs redirect the slowed atomic beam into the main chamber. The strontium oven and optics for the 2D MOT and Zeeman slower are commercial from AOSense Incorporated.

Atoms are then captured in the main chamber with a three dimensional ``blue'' MOT, named for the color of the trapping light at $\approx$461~nm. Spatially dependent magnetic field gradients along each axis impart an energy shift of the excited state Zeeman sublevels dependent on the position of the atoms from the center of the trap. Counter propagating beams with polarizations set to $\sigma^{+}$ and $\sigma^{-}$ are once again detuned at $\Delta/2\pi = $-40~MHz from the ${}^1\!S_0 - {}^1\!P_1$ transition and interact with the off center atoms further cooling while imparting a restoring force for trapping. 

During the ``blue'' MOT stage, about 1:50,000 atoms in the $^1\!P_1$ state decay to the $5s5p{}^3\!P_2$ metastable reservoir state. A laser beam tuned to the $5s5p\;^3\!P_2 - 5p^2$\;$^3\!P_2$ transition at 481~nm, re-pumps the atoms from the metastable state to the $^3\!P_1$ where they are further cooled and trapped in the second stage, ``red'' MOT, operating on the $^1\!S_0 - ^3\!P_1$ transition at $\approx$689~nm using an external cavity diode laser with a linewidth below 1~kHz.

The ``red'' MOT stage operates in two phases. The sawtooth wave adiabatic passage (SWAP) technique \cite{Bartolotta2018, Snigirev2019} is applied during an initial broadband phase wherein the laser detuning is swept at a rate of 40~kHz in a triangle waveform from $\Delta_{-}/2\pi=-10$~MHz to $\Delta_{+}/2\pi=+200$~kHz followed by a lower power, single frequency phase with the detuning fixed at $\Delta/2\pi=-290$~kHz. At the end of both phases, we capture $5\times 10^{5}$ atoms on average.

The temperatures achieved after each MOT stage are limited by the natural linewidth of the relevant transition. The ``blue'' MOT ($\Gamma/2\pi$=30.5~MHz) reaches temperatures on the order of 1~mK and the ``red'' MOT ($\Gamma/2\pi$=7.6~kHz) achieves temperatures of $\approx$1~$\mu$K.

During the ``red'' MOT cooling and trapping, the ODT beam is overlapped with the atomic cloud for 100~ms before turning off the MOT beams for 10~ms and taking an absorption image of the atoms. The imaging probe beam operates on the narrow-line, 689~nm transition with $I/I_{\rm{sat}}=0.15$ and pulse length of 50~us. The absorption imaging light is collected on an Andor 897 EMCCD camera.

The ODT beam is aligned across the center of the ``red'' MOT by setting the ODT beam frequency at the resonance frequency of the $5s5p\;^3\!P_1 - 5p^2\;^3\!P_2$ transition, determined from the literature to be at $472.36$~nm. The ODT beam induces a large AC Stark shift on the atoms rendering them transparent to the MOT cooling beams. As the resonant ODT beam is aligned towards the center of the cold atomic cloud, the number of atoms is minimized. We repeatedly decrease the power and continue alignment until only the atoms in the center of the atomic cloud interact with the ODT beam. When the beam is aligned, a dimple appears in the center of the atomic cloud where no atoms are trapped. 

\subsection{Spectroscopy}

With the fitted beam waists of $44$ and $58$~$\mu$m at the atoms for $m=0$ and $m=-1$ measurements, we calculate the Rayleigh range to be $13$~mm and $22$~mm. The atoms are trapped in the ODT in the shape of a narrow cylinder whose length, while less than the Rayleigh range, is much longer than the width. The probe beam being colinear with the ODT beam images the trapped atoms along the beam direction at high optical density. The high density atomic cloud induces a lensing effect in the absorption images when the probe beam detuning is $\approx\pm50$~kHz from resonance. These images exhibit constructive interference of the imaging light in different regions of the image for detunings $+50$~kHz and $-50$~kHz from resonance creating unsymmetric spectroscopy scans. This is solved by adding a linear model to the Voigt profile for the spectroscopy fits.

\subsection{Fitting procedure}
In Fig. \ref{fig:results}, all of the black lines are related to the data collected. We fit a general quadratic model ($\alpha\lambda^{2} + \beta\lambda + \gamma$) using a weighted least squares fitting procedure with the weights given by $1/DSSC_{err}$ where $DSSC_{err}$ is the error of each DSSC measurement. The 95\% confidence interval is calculated using the error of the fit and a two-sided student-t value for 3 degrees of freedom ($\approx$3). The 95\% prediction interval incorporates both the error of the fit and the error of the DSSC measurements using the same student-t value. We then use the 95\% prediction interval to perform a Monte Carlo root finding method of the quadratic fit parameters with 10000 iterations. The magic wavelength is given by the mean of all 10000 iterations and the error bar is given by the standard deviation. Thus the 95\% prediction interval will be $\approx$3 times greater than the magic wavelength errorbar.

The red theory curve is fit to the six data points leaving only the waist as a free parameter. The dashed-dot lines correspond to $\delta\alpha$ from the theoretical calculations plotted in Fig. 1.

\subsection{Noise sources}

The dominant noise source for the $m=0$ measurements is the laser power variation of the ODT beam, with power fluctuations below $2$~\% over the time it takes to complete a single spectroscopy frequency scan. The $m=-1$ measurements are in addition affected by variations in intensity of the bias and background magnetic fields.

The electrical current generating the bias magnetic field is stabilized with a PID feedback controller and is stable for low frequency noise up to 100~Hz. The entirety of the feedback system generates high frequency peak-to-peak noise levels of 5~mV which, through the current transducer calibration and the magnetic coil setup, corresponds to a worst case fluctuation of $\pm5$~mG at frequencies greater than 100~Hz. Given the $2.1$~kHz/mG Zeeman splitting under the magnetic field bias, this generates a large noise source of $\pm 10.5$~kHz which is evident in the noisier $m=-1$ DSSC data. The background field stability is below the measurable resolution of our external magnetometer of $10$~mG, an AlphaLab VGM. Any long term drift of the magnetic field is accounted for by taking a second reference scan as mentioned in the text. 

Our High-Finesse WSU-2 wavemeter is calibrated daily with our kHz-linewidth $\approx$689~nm laser locked to the ${}^1\!S_0 - {}^3\!P_1$ transition. At 473~nm, the ODT wavelength is beyond the calibration specification of the wavemeter and thus induces a systematic offset in the wavelength reading. To measure this offset, we calibrate the wavemeter to the ${}^1\!S_0 - {}^1\!P_1$ transition at $\approx$461~nm and set our ODT wavelength to 473.000000~nm. Then, we re-calibrate it back to the 689~nm laser and note the ODT wavelength becomes 472.999600~nm. Thus, we add an additional +0.4~pm offset to our measured ODT wavelengths for both the $m=0$ and $m=-1$ data, which is an order of magnitude smaller than the smallest reported uncertainty of 4~pm. 

\subsection{$\theta_{p}$ Measurement}
Any deviation of $\theta_{p}$ from $0^{\circ}$ causes a systematic error as well as further uncertainty in the magic wavelength measurement. We measure the angle between the ODT polarization and the $\vec{B}_{bias}$ field using an indirect measurement \cite{Zheng2020}. Since the probe beam and the ODT beam exist in the same plane, we ensure both beams are aligned to the same axis, limited only by our measured extinction ratio of 18~dB. If the probe beam polarization is aligned to ${\vec{B}_{bias}}$ ($\theta_{p} = 0^{\circ}$), there will only be excitation of the $\pi$ transition. Alternatively, if the probe beam polarization is orthogonal to ${\vec{B}_{bias}}$ ($\theta_{p} = 90^{\circ}$), there will only be excitation of the $\sigma^{-}$ transition. Any deviation from $\theta_{p}=0^{\circ}$ will cause a non-zero population in the $\sigma^{-}$ excitation.

The excited state populations are related to the Rabi frequency of each by $N_{\pi} = N_{g} \sin^{2}(\frac{t\Omega_{\pi}}{2})$ and similarly for $\sigma^{-}$. Expanding the $\sin()$ terms yields a relationship between the population ratio and the ratio of Rabi frequencies. We deduce the angle offset by $\theta_{p} = \arctan(\frac{\Omega_{\sigma^{-}}}{\Omega_{\pi}})$

\end{appendix}

\bibliographystyle{apsrev4-1}
\bibliography{dssc_bib}

\end{document}